\documentstyle[aps]{revtex}

\begin{document}

\title{Metallization of Fluid Hydrogen}
\author{W.J. Nellis}
\address{Lawrence Livermore Laboratory
University of California
Livermore, California 94550}
\author{A. A. Louis and N.W. Ashcroft}
\address{Laboratory of Atomic and Solid State Physics,
Cornell University, Ithaca, NY 14853-2501}

\maketitle
\begin{center}
\end{center}

\begin{abstract}

The electrical resistivity of liquid hydrogen has been measured at the high
dynamic pressures, densities and temperatures that can be achieved with a
reverberating shock wave.  The resulting data are most naturally interpreted
in terms of a continuous transition from a semiconducting to a metallic,
largely diatomic fluid, the latter at 140 GPa, (ninefold compression) and
3000 K.  While the fluid at these conditions resembles common liquid metals
by the scale of its resistivity of 500 micro-ohm-cm, it differs by retaining
a strong pairing character, and the precise mechanism by which a metallic
state might be attained is still a matter of debate.  Some evident
possibilities include (i) physics of a largely one-body character, such as a
band-overlap transition, (ii) physics of a strong-coupling or many-body
character,such as a Mott-Hubbard transition, and (iii) processes in which
structural changes are paramount.

\end{abstract}

\vspace{1cm}

\newpage

%\narrowtext 
%\twocolumn

\section{Introduction}

	Hydrogen has been cited as the prototypical system for the study of
the insulator-to-metal (IM) transition since Wigner and Huntington predicted
in 1935 that the insulating molecular solid could transform to a conducting
monatomic solid at sufficiently high pressure at $0 K$\cite{Wign35}. That
is, although solid molecular hydrogen is a wide bandgap insulator ($E_g$ =
15 eV) at ambient conditions, at sufficiently high pressure the insulating
diatomic solid is expected to transform to a conducting monatomic solid or
the electronic energy bandgap $E_g$ of the diatomic solid is expected to
close, resulting in an IM transition.  Since Wigner and Huntington's paper,
predictions of the transition pressure of the former has varied from
25\cite{Wign35} to 2,000 GPa\cite{Alde60} at $0 K$.  To date this transition
in the solid at low temperatures has not been observed by optical
measurements
 in the range 190 to 260 GPa.\cite{Chen96,Heml96,Ruof96}.
	
Metallization within the molecular solid phase by a band overlap mechanism is
predicted to occur at pressures lower than is the case for the transition to
the monatomic phase\cite{Frie77}, but the transition pressure is
structure-dependent\cite{Garc90,Kaxi91,Chac92}, and the structure at
densities close to metallization at 0 K is not known.  Extrapolation of
recent pressure-volume data up to 120 GPa in the hcp phase yields a
predicted dissociative transition of 620 GPa\cite{Loub96}.
 Both the monatomic and diatomic metallic solids have been predicted to be
high-temperature superconductors\cite{Ashc68,Rich97}.

	Electrical conductivity measurements indicate that hydrogen becomes
metallic (i.e. conducting\cite{metallic}) at 140 GPa, ninefold compression
of initial liquid density, and 3000 K\cite{Weir96}. Metallization is expected
when pressure and density are sufficiently high that the electronic bandgap
decreases from the value at ambient conditions of 15 eV down to $\sim 0.25$
eV, the temperature of these experiments.  Extrapolation by means of the
Simon equation of the melting curve of $H_2$ measured at low
pressures\cite{Diat85} gives a melting temperature of 1800 K at 140 GPa.
Thus, as stated,  metallization probably occurs in the high temperature fluid.

	Electrical conductivity has also been measured under single-shock
compressions up to 20 GPa and 4600 K\cite{Nell92}. Those measurements showed
that electronic conduction is thermally activated in the semiconducting
fluid.  Electrical conductivity experiments using explosively driven
magnetic flux compression to isentropically compress liquid hydrogen have
shown that the conductivity becomes greater than $1 (\Omega cm)^{-1}$ at 200
GPa and $400$ K\cite{Hawk78}.  Although this conductivity does not conform
to metallic values, this experiment demonstrates that the electrical
conductivity increases at high pressure and temperature.  Shock Hugoniot and
temperature data have also been measured\cite{Nell83,Holm95,Silv97}. 

     Hydrogen is important for astronomy because its cosmological abundance
is about 90 atomic percent, and the understanding of dense hydrogen is
particularly important to planetary science. 
 Jupiter and Saturn contain over 400 Earth masses, most of which is fluid
hydrogen.  Jupiter-size planets now being discovered close to nearby
stars\cite{Butl96} probably contain massive amounts of hydrogen as well, and
the interiors of these giant planets are likely to be at high pressures and
high temperatures and in the fluid state\cite{Zhar92}. Because of the large
mass diffusion coefficient and low thermal conductivity\cite{Ross81},
magnetic fields are produced by the convective motion of electrically
conducting fluid hydrogen by dynamo action\cite{Stev83}. Implications for
Jupiter of recent measurements on dense hydrogen at high pressures and
temperatures have been described elsewhere\cite{Nell95}.

	Because of the high kinetic energy in the impactor (0.5 MJ), in a
sense we are actually working at a confluence of High Energy and Condensed
Matter Physics.  This energy is comparable to the total kinetic energy of
all the protons and antiprotons in the beams in the Tevetron at the Fermi
National Accelerator Laboratory.  Here energy enables discovery of novel
states of condensed matter, analogous to the discovery of novel states of
subnuclear matter.

\section{  Finite Temperatures}

	The distinguishing feature of these experiments is achievement of a
stable hydrogen sample at 3000 K and 100 GPa pressures.  It is desirable to
look for the IM transition at higher temperatures because phenomena which
inhibit metallization in the dense solid; namely, crystalline and
orientational phase transitions\cite{Ashc95,Edwa97}, do not occur in the
disordered fluid.  However, because of the large mass-diffusion coefficient
and chemical reactivity of hydrogen at high temperatures, it is essential to
use a temperature pulse.  The duration of this pulse should be sufficiently
long to achieve equilibrium and sufficiently short that the sample cannot
diffuse away nor react chemically before the experiment is completed.  The
$\sim 100$ ns duration of shock compression satisfies these criteria.  Our
temperature (equivalent to 0.3 eV) is relatively low for the electron
distribution being probed because the energy gap at ambient pressure is 15 eV
and the zero point energy of the molecule is 0.3 eV.

\section{  Experiment}

	Conditions of high pressures, densities, and temperatures were
produced by impact of a planar metal plate onto a cryostat containing liquid
hydrogen.  A two-stage light-gas gun was used to accelerate the impactor
plate up to $\sim 7 km/s$\cite{Jone66}.  Hydrogen gas is used to accelerate
the projectile because it has the highest sound speed of any gas and thus
produces the highest impact velocity and pressure.  The magnitude of the
pressure generated by the impact is determined by the impact
velocity\cite{Mitc81a} and the Hugoniot equations of state of impactor and
target\cite{Mitc81b,Ersk94}.  To achieve highest densities and lowest
temperatures at high pressures, the sample must have a relatively high
initial density.  High densities were achieved by using a liquid sample at
20 K.\cite{Nell83,Nell80}.  We applied our previous techniques to measure
electrical conductivities\cite{Nell92,Rado86}
 to the
configuration illustrated in Fig. (1).

	The sample is 0.5 mm thick and 25 mm in diameter contained between
two electrically insulating sapphire (single-crystal z-cut $Al_2O_3$) anvils,
2.0 mm thick and 25 mm in diameter.  These are contained between two 2.0
mm-thick Al disks.  Al is strong, ductile, and a good thermal conductor at
20 K, which facilitates condensing the sample from high-purity gas.  Both
$H_2$ and $D_2$ samples were used, depending on the final density and
temperature desired.  At the relatively high final temperatures achieved
($\sim 3000$ K), possible effects of different zero-point energies for hydrogen
and deuterium are negligible.  Stainless steel electrodes exit the sample
holder to the right.  In the metallic phase four electrodes are used; two
outer electrodes for current and two inner electrodes to measure potential
difference.  Two electrodes are used if hydrogen is in a state of poor
 conductivity.  The
electrical circuit for four-probe measurements is shown in Fig. (1b).  Shock
compression switches the hydrogen sample into the circuit when its
conductivity becomes significant.  Voltages were measured differentially
with 1-ns time resolution.  Current was measured with a Rowgowski coil.
Typical currents were $\sim 1$ A and voltages were a few tens of mV.
 The resistance $R$ is directly proportional to the resistivity $\rho$, so
that $R = C \rho$.  To determine the cell constant C, steady-state,
three-dimensional current flow simulations were performed to calculate $R$ and
$\rho$\cite{Weir96}.

\section{  Thermodynamic States}

	This experiment achieves highly condensed matter relatively close to
the 0 K isotherm by means of a reverberating shock\cite{Yoo89}. The impact
generates an initial pressure $P_f$ in the $Al_2O_3$.  When this shock
reaches the liquid hydrogen, the pressure of the shock drops until the
release pressure of sapphire matches the Hugoniot of liquid hydrogen.  This
drop is about a factor of 30 in pressure.  The shock in hydrogen then
reverberates back and forth between the sapphire anvils until the pressure
reaches $P_f$, the pressure incident initially from the sapphire.  Thus, the
first wave in hydrogen is a weak shock and the total of the successive shock
reverberations is a quasi-isentrope.  This fast compression is near
adiabatic and the temperature of hydrogen rises.  This process causes a final
temperature which is about an order of magnitude smaller than would be
achieved by a single shock to the same final pressure.

	The initial weak shock occurs in $<10^{-12} s$ (Fig.(2a)); the following
quasi-isentrope takes $\sim 5\times10^{-8} s$.  The corresponding effect on
the thermodynamics is illustrated in Fig.(2b).  These various pressure-density
states were calculated using the equation of state of Kerley\cite{Kerl83}.
Shock reverberation is necessary to reach metallization near $0.7 g/cm^3$.

	The densities and temperatures must be known to analyze the
electrical conductivities.  At present no means to measure them exist; they
must be calculated.  The densities and temperatures were calculated with two
 equations of state for hydrogen.  One was developed by Kerley\cite{Kerl83}
before recent shock data were available; it neglects molecular dissociation
at the conditions in our experiments.  The other is due to Ross, which is
based on our recent shock temperature data and includes molecular
dissociation\cite{Holm95}. The preferred values were those calculated with
Ross' model, because it is based on recent data.  The calculated final
pressures in the hydrogen or deuterium agree to within 1\% with the initial
shock pressures in the sapphire calculated by shock impedance matching.
Based on these calculations, the systematic uncertainties in calculated
density and temperature are 5\% and 20\%, respectively.

\section{  Analysis of Conductivity Data}

	The experimental resistivity data are plotted as $\log(\rho)$ versus
pressure in Fig. (3).  The change in slope at 140 GPa is indicative of the
transition to the metallic state.  We analyzed results in the semiconducting
range, 93-135 GPa, by fitting the data to the dependence of conductivity for
a thermally activated semiconductor:
\begin{equation}\label{Billeq1}
   \sigma = \sigma_0 \exp (- E_g(D) /2k_BT),
\end{equation}
where $\sigma$ is electrical conductivity, $\sigma_0$ depends on density
$D$  and relatively weakly on T, $E_g(D)$ is the density-dependent mobility
gap in the electronic density of states of the fluid, $k_B$ is Boltzmann's
constant, and $T$ is temperature.

	Seven data points with error bars of 20 to 50\% were fit to Eq.
(1),  the results being: $E_g(D) = 1.12 - ( 54.7 )( D - 0.30 )$, where
$ E_g(D)$ is
in eV, $D$ is in $mol/cm^3$ $(0.29-0.32$ $ mol/cm^3)$, and $\sigma_0 =
 66$ $(\Omega cm)^{-1}$.  Similar results are obtained using densities and
temperatures calculated with Kerley's model\cite{Weir96}. $\sigma_0 =
200-300$ $(\Omega cm)^{-1}$ is typical of liquid
semiconductors\cite{Mott71a}.  Since $\sigma_0$ is within a factor of 3 of
the typical value, this result is reasonable.  The theoretical rate of
bandgap closure near metallization at 0 K of molecularly disordered hcp
hydrogen is $40 eV/(mol/cm^3)$\cite{Chac92}, which is comparable to our
value of $\sim 60 eV/(mol/cm^3)$ obtained by fitting.  Thus, this slope is
also comparable to what is expected.  From our fit the gap $E_g(D)$ and
$k_BT$ are equal at a density of $0.32$ $mol/cm^3$ and a temperature of
$\sim 2600 K$ $(0.22 eV)$.  In this region the energy gap is smeared out
thermally, activation of electron carriers is complete, disorder is
saturated in the fluid, and conductivity is expected to be weakly sensitive
to further increases in pressure and temperature, provided the fluid does
not change significantly.  At $0.32 mol/cm^3$ the pressure is 120 GPa, which
is close to the 140 GPa pressure at which the slope changes in the
electrical resistivity (Fig. (3)).

	At pressures of 140 to 180 GPa the resistivity changes character; it
is essentially constant at $500 \mu \Omega cm$, equivalent to a conductivity
of $2000 (\Omega cm)^{-1}$.  This value is typical of the fluid (monatomic)
 alkali metals Cs and Rb at 2000 K undergoing the same
transition\cite{Hens96}.  Also, the minimum electrical conductivity of a
metal is given by $\sigma =2\pi e^2/3 h a$, where $e$ is the charge of an
electron, $h$ is Planck's constant, and $a$ is the average distance between
particles\cite{Mott71b}.  In this case $a \sim D_m^{-1/3}$, where $D_m$ is the
density of hydrogen at metallization.  The calculated minimum metallic
conductivity is $4000 (\Omega cm)^{-1}$, which is in good agreement with the
experimental value of $2000 (\Omega cm)^{-1}$.  Thus, fluid hydrogen becomes
conducting at about 140 GPa and 3000 K via a continuous transition from a
semiconducting to metallic fluid, in which the electronic activation energy
is reduced by pressure to $k_BT$.

An important point to keep in mind in what follows is this; though the
characteristic time of the experiments are probably long enough to achieve
equilibrium in a single phase, it is possible that the system has not
reached its equilibrium phase should it happen that the system has just
crossed a phase boundary.

\section{Theoretical Considerations}

In hydrogen, at the conditions of the experiment, the confluence of thermal
and quantum effects link the physics of dissociation and metallization,
electronic and atomic ordering, and ionization and chemical (cluster)
formation.  Together these can lead to a remarkably rich phase diagram, most
of which remains unexplored.  The metallization experiment\cite{Weir96}
reveals a new section of this phase diagram, and gives hints to its further
structure, as shown in Fig. (\ref{phasediagram}). Traditionally, theoretical
interest has focussed on the change from a diatomic to a monatomic liquid or
plasma, as well as the change from a metal to an insulator.  In the new
experimental regime now opened up, translational energies appear to match
the vibron energies, and as such the metallization experiment may be probing
both types of changes.  With these confluences in mind we comment in turn on
some interesting theoretical questions raised by the experiment, namely:
(a) Why does the resistivity, the primary measured quantity, take the value
it does?, (b) Why does the metal-insulator transition occur at a considerably
lower pressure in the liquid than is predicted in the solid?, (c) What is
the detailed mechanism of the metal-insulator transition?, and (d) What
further experimental probes might clarify the situation?

\section{The scale of resistivity in liquid metallic hydrogen}

The fundamental approach to the transport properties of liquid metals is via
the Kubo-Peierls-Greenwood theory, based on linear response of the system to
a weak external field $E(t)$.  The static conductivity $\sigma(0)$ is
then given by:
\begin{equation}\label{eq1aa}
\sigma(0) \sim \int_0^\infty  <j(0)j(t)>dt
\end{equation} where $<j(0)j(t)>$ is a current-current correlation function
for the many-electron system.  If the self-energy of the electron system is
sufficiently small (from all sources of interaction) then an expansion in
the self-energy leads to an inversion of $\sigma$ for the resistivity, and
at lowest order this is equivalent to first-Born approximation within the
Boltzmann semi-classical approach to transport.  In the high temperature
limit it leads to the well known Ziman formalism~\cite{Zima61,Fabe72,Shim77}
for degenerate electron systems, and this has been remarkably successful at
predicting the resistivity of liquid metals to within about a factor of
two, as well as trends seen in alloys, thermopower and the temperature
dependence of the resistivity\cite{resliquid}.  It requires as input the
static structure factor $S(q)$ of the scattering system, an
electron-ion-interaction, and the assumption of an
appropriate high temperature limit.

  To assess the applicability of this approach to the present problem, we
may consider the case of liquid Si and Ge which prior to melting are not
metals, though conduction by thermally excited carriers is evident.  As
fluids they are significantly under-coordinated ($\sim 5-6$) which reflects
persistent covalent effects and hence interactions going beyond the pair
model.  In applications of the Ziman formalism to this problem, the tangible
effects of remnant covalency are included only through the static structure
factor $S(q)$; 4 electrons per ion are usually assigned to the conduction
process [but see below].

In some ways, the situation for hydrogen can be considered physically
similar to Si or Ge, only more extreme.  It is, again, a fundamentally
covalent (but not necessarily a network) system, and structurally it cannot
be described just in terms of pairwise proton-proton interactions.
 The viewpoint taken from liquid Si (a metal) can be taken here if the state
possesses a Fermi surface:  in the
application of the weak scattering formalisms covalency manifests primarily
through $S(q)$ and in the specification of the fundamental excitations
associated with the structure.  As a first approximation (compare with the
case for Si), two electrons per ``molecule'' could be taken to participate in
conduction, but with a significant density of states correction to be
applied later.
 Given this fundamental picture essentially tied to the pairing picture of
the solid, we proceed to investigate the resistivity of liquid metallic {\em
molecular} $H_2$ (in the experimental metallization regime) starting with the
 Ziman expression~\cite{Zima61,Fabe72,Shim77}.  We express the
resistivity $\rho_L$ originating with  a given type of carrier as:
\begin{equation}\label{eq1a}
\rho_L = \frac{m}{n_c e^2 \tau},
\end{equation} where $m$ is the carrier mass, $n_c$ is the carrier density,
and $\tau$ is a measure of the relaxation time.
In its simplest form the Ziman formalism
gives\cite{Loui97}:
\begin{equation}\label{eq3}
\rho_L = \left( \frac{a_0 \hbar}{e^2} \right) \frac{6 \pi}{(n_s)(k_F)^2}
 \int_0^1 S(y)|v(y)|^2 (y)^3
 d(y),
\end{equation}
 where $y = q/2k_F$ is the wave vector scaled by twice the Fermi wave-vector,
$n_s$ the number density of scatterers, $v(y)$ the screened coulomb
potential, $S(y)$ the
 structure factor describing the proton-proton density fluctuations 
which scatter the carriers, and $(a_0 \hbar/e^2)$ can be viewed as the atomic
unit of resistivity (it has the value of $21.7 \mu
\Omega cm$).  

 The structural average required in the static structure factor appearing in
(\ref{eq3}) is not simple given the conditions of the experiment.  In the
first place translational energies appear well matched to rotational
energies, and also to vibron energies.  The kinematic transfer of energy
from translational to vibrational degrees of freedom will be competitive
with transference of translational energy to rotational energies.
Accordingly, for this system the temperature and the density conspire in
such a way that a standard averaging over rotational degrees of freedom
prior to translational averaging is not secure.  Secondly, orientational and
translational entanglement leads to considerable self-averaging in the
determination of the static structure factor.  Thirdly, it has to be
emphasized that so far as the vibron excitations are concerned, the system
is not at high temperatures, which affects both structural\cite{struckcorr}
and scattering\cite{scattcorr} physics.

In a naive approach using the usual free-electron density of states, and
 with the approximations introduced above for the structure factor $S(q)$, a
resistivity of about
 {\bf $50$ $\mu\Omega cm$} is found from (\ref{eq3}), a significantly larger
figure than obtained in previous calculations of monatomic H\cite{Stev74},
but still considerably lower than the $500$ $\mu \Omega cm$ measured
experimentally\cite{Weir96}.  However, this is to be expected since in the
one-electron picture we also anticipate  a significant decline in
the density of states at the Fermi energy\cite{noEdwards}.

If the band-gap has only just closed we expect a lowering of the density of
states as compared with the full free-electron case which in turn can be
linked to a significantly reduced effective carrier density. 
 Equation~(\ref{eq1a}) shows that a reduced number of carriers raises the
resistivity\cite{meanfreepath}; we also expect related multiple-scattering
and other strong-coupling effects\cite{Kita95} to be relatively important in
this regime, and these effects also lead to higher resistivities.  Thus by
taking into account physics beyond the usual one-band picture our 
result of $\sim 50 \mu \Omega cm$ can only be seen as something of a lower
bound.  Since the overlap and related electron density is expected to
increase with increasing density, this modified Ziman picture predicts that
the resistivity at fixed temperature should continue to fall with increasing
pressure so long as the liquid maintains its pairing correlations.  We
emphasize, however, that the pairs are in no sense permanent (again an
analogy to fast exchange in liquid Si can be made) although they are
long-lived on {\em electronic} timescales and as such influence the 
electronic structure.

At present the experimental results have large error bars ($20 \%$ in the
metallic state) and are in too small a pressure range to confirm or falsify
any predicted pressure dependence of the resistivity, but the actual value
of the resistivity near metallization seems consistent with this picture,
though this is not the only possible interpretation (see below).

\section{The onset of the insulator to metal transition}

At first sight the lower pressure and density of the metal-insulator
transition in the liquid compared to the solid may seem surprising.  One
would expect the greater disorder (as compared with the solid) to delay the
closing of the gap and the subsequent onset of metallization as is for
example the case in Hg, another divalent pressure-induced metal-insulator
transition\cite{Kres97}.  However, a closer look at the origin of the
bandgap in the solid phase reveals a more complex possibility which can be
illustrated in the Pa3 structure, an isotropic solid which best mimics the
high coordination and orientational averaging characteristics of the
``paired'' liquid.  The lower bands in such a structure can be represented by an
equivalent face-centered cubic description\cite{Ashc95b} where the principal
Fourier components of the electron interaction with pairs are given by
\begin{equation}\label{eq4}
V_{111} = (3/4 \pi r_s^3 a_0^3)S_0(111)v(111)
\end{equation}
and 
\begin{equation}\label{eq5}
V_{200}  =  (3/4 \pi r_s^3 a_0^3)S_0(200)v(200).
\end{equation}
   Here $S_0$ is the structure factor per proton, and is given by
\begin{eqnarray}\label{eq6}
 S_0(l,m,n) =  \frac{1}{4} \{  \cos 2 \pi
\alpha(l+m+n)
 &+& (-1)^{m+n}\cos 2
\pi \alpha(l - m + n) \nonumber \\ 
 &+&  (-1)^{l+m}\cos 2 \pi \alpha(-l + m + n) \nonumber  \\
 &+& (-1)^{n+l}\cos 2\pi \alpha(l + m - n)\}
\end{eqnarray} with $\alpha=d(r_s)/\sqrt{3a}$ ($2a$ being the separation, on
average, between protons), and $v(K)$ is the Fourier transform for the
screened potential of a single proton.  

Increase of the temperature to ($ k_B T \sim \hbar\omega_{vib}$)
  significantly excites the pairs into higher anharmonic vibrational states
resulting in large amplitude spatial fluctuations. These in turn will lower
the gap by a corresponding Debye-Waller factor, and given that the
excursions are appreciable, the lowering can be significant.  This can be
demonstrated most easily by extending one of the 4 pairs in the Pa3
structure to a separation of $2d'$ ($2d' > 2d$) while the other basis pairs
are kept fixed.  From (\ref{eq6}) the structure factor per proton for such a
separation becomes
\begin{equation}\label{eq8}
S(l,m,n) = S_0(l,m,n) - \frac{1}{2} \sin[ \pi (\alpha' - \alpha)(l+m+n)]
\sin[\pi(\alpha + \alpha')(l+m+n)],
\end{equation} which shows immediately that the gap will be reduced when the
pair extends.  This gives a strong hint of the likely effect of temperature,
primarily to reduce the gap, and to promote an earlier transition from
insulator (or semi-conductor) to metal.  Note  that a displacement
arising with acoustic phonons can also contribute to this effect.

\section{The Physical Characteristics of the Insulator -Metal transition}

At the one-body level the general closing of the gap with increasing
pressure and density appears well established, but what exactly happens upon
close approach to the metal-insulator transition in hydrogen?  Some
possibilities now follow:

\subsection{Scenario 1: Metal-insulator transition induced by one-body physics}

In a continuous single particle approach, the gap closes and progressively
more and more carriers are excited into the conduction band; structurally
this implies strong pairing correlations.  Depending on the decline of this
gap, the emerging conductivity can be significant.  Experiments and theory
converge on a consistent picture for hydrogen where the
 band-gap decreases with increasing pressure, both in the
solid\cite{Chac92,Mao94} and liquid\cite{Nell92} phases.  In the solid, the
concept of an indirect gap closing with increase of density in the paired
phase is robust when calculated in various crystal structures and it also
obtains under orientational disorder\cite{Frie77,Kaxi91,Chac92}; if
generalized to the separation between mobility edges\cite{Mott90}, we expect
it to carry through for the liquid phase as well\cite{Bismuth}.
Accordingly, under the assumption that the liquid remains largely paired,
this view leads to a picture where the one-electron physics of the liquid
may be similar to that of the solid; under pressure mobility edges are
gradually approaching, and if this continues hydrogen will be transformed
from a liquid semi-conductor to a liquid semi-metal.
 As pressure increases further the ``bands'' overlap (mobility edges cross),
and the effective carrier density continues to increase, eventually becoming
degenerate\cite{bands}. 

The vibrational dynamics of the $H_2$ molecules imply an interesting
possibility for the scale of conductivity in the semi-metallic regime (i.e.
$k_BT
\sim E_g$).  If we assume that the electrons hop from pair to pair when
close approach sufficiently increases orbital overlap, then their time-scale
will be set by those translational excursions 
 which bring molecules into sufficient proximity that wave-function overlap
is especially effective (i.e. $\tau
\sim 10^{-14}$ s).  Together with a standard semi-conductor picture
invoked to estimate the carrier density (see e.g. p 575 of 
\cite{Ashc76}), this quickly leads to resistivities as low as $\sim 700 \mu
\Omega cm$ for $E_g \sim 2k_B T$\cite{Loui97}!

Note, however, that the liquid phase corresponds to proton arrangements that
are strongly disordered.  If viewed at low temperatures, disorder can lead
to states that lack diffusion, via Anderson-localization.  The role of
temperature is then to excite electrons from such states.  Note however,
that the condition for Anderson localization is similar to the condition for
the Mott transition:
\begin{equation}\label{eq9}
D_m^{1/3} a^* \sim C,
\end{equation} where $a^*$ is a measure of the localized orbital size, and C
is typically about $0.3$\cite{Mott90}. Under the assumption that the Bohr
radius for diatomic hydrogen is roughly the same as the Bohr radius of
monatomic hydrogen,  $D_m^{1/3} a^* \sim 0.3$ at experimental
metallization conditions.

\subsection{ Scenario 2: Metal-insulator transition and the role of many-body
physics}

 Mott\cite{Mott90} argued that a gap will never close continuously because
upon restoration of residual many-body interactions, the formation of
excitons which subsequently unbind upon further closing of the gap will lead
to a jump in the conductivity.  We expect in this case that the temperature
of $\sim 0.3$ eV will be too high for the formation of excitons\cite{Halp67}
but this does not rule out some other type of correlation driven transition
near band-gap closure. For example, the Hubbard $U$ for monatomic hydrogen is
about 17 eV\cite{Mott90} while the free-electron band-width at experimental
conditions is around 20 eV, which is probably an upper bound. Since the
effects of disorder and correlation are expected to reinforce each
other\cite{Thou79} the possibility of conductivity by thermal excitation
across a declining Mott-Hubbard gap emerges. If we are in the regime where
electron-electron interactions play an important role, the effect of spin
becomes non-trivial. 
  The twin effects of strong correlation and disorder as they influence the
metal-insulator transition are not completely understood; they combine with
some force in the hydrogen problem.
  
The highly successful semi-empirical Herzfeld polarization catastrophe
criterion predicts that $H_2$ will become metallic only upon about $\sim 11$
fold compression\cite{Edwa83} (generally for ordered phases). 
 However, we note that there is some ambiguity in the definition of the
various quantities entering these criteria (for instance the meaning of
local polarizability in the present environment), a problem that becomes
even more acute for hydrogen where the fluid phase is involved.

\subsection{ Scenario 3: Metal-insulator transition induced by structural
change}

If simple pressure induced band-gap closure in the paired state is preceded
by significant structural change, metallization can also precede complete
closure of this gap.
 Perhaps the best known case of a metal-insulator transition driven by
structural change is the metallic behavior seen when melting the elemental
semi-conductors such as Si and Ge (as discussed above).  The change in
band-structure is driven by the change in local atomic structure, in this
case a relatively small increase in coordination.  In the case of hydrogen,
structural change associated with, for example the rapid decline of pairing
correlation, is a likely candidate.  In the liquid phase, we have a  
direct parallel of the crystalline case discussed in section VIII. above.

In fact, the Wigner-Huntington proposal\cite{Wign35} itself falls into the
class of MI transitions induced by structural changes. Here it was predicted
that increasing pressure (at T=0) could lead to dissociation in hydrogen,
resulting in a structure with one electron per unit cell and concomitant
metallic behavior for paramagnetic states\cite{WignerMott}.  Even at
relatively low densities, increasing the temperature to regimes comparable
with the dissociation energy will also rupture the H-H bond, possibly
leading to metallic (conducting) behavior.  The case at hand is clearly much
more subtle.  At the extreme pressures of the shock experiment, the H-H bond
is significantly weakened, but even at $r_s=1.5$ ($\sim$ 9 fold compression)
the corresponding well depth continues to have a magnitude significantly
larger than the temperatures achieved\cite{well}.  To date no evidence of
static pressure induced dissociation in the solid has been observed, even at
the significantly higher pressures of $250 GPa$ achieved by ultra pressure
diamond anvil-cell techniques.  Although the evidence above refers to
zero-temperature solids in particular structures, increasing the temperature
is not likely to change the well depth significantly.  

At these extremely high densities and temperatures, the very concept of {\em
dissociation} is nebulous.  High temperatures will excite large amplitude
stretching motion, leading for example to exchange of atoms from neighboring
pairs, or to the transient formation new localized species (``$H_3$'' etc...)
and so standard chemical definitions are not unique.  Perhaps it is better
to speak of local coordination in a completely statistical sense, for
example as the integral of the pair correlation function carried out to a
certain distance.  These changes in local coordination certainly lead to
corresponding changes in the global electronic structure as pointed out in
an interesting recent tight-binding (TB) molecular dynamics study by Lenosky
et al.\cite{Leno97}.  Here ``dissociation'' is defined by asserting that two
atoms constitute a dimer if each forms the nearest neighbor atom of the
other.  Atoms not identified as belonging to a dimer are then termed
dissociated ``monomers''.  Interestingly, they find that states near the
chemical potential have a large projection onto these ``monomers'', and so by
virtue of the Mott formula, for which resistivity scales as the inverse
square of the density of states at the Fermi energy, the resistivity in
their simulation scales inversely with the square of the average number of
``monomers''.  The mechanism they propose (similar to that suggested by
Ross\cite{Ross96}) in which the ``dissociation'' plays a crucial role is a
possible candidate for what might happen in liquid metallic hydrogen.
However the large Hubbard U, included only indirectly through the effective
parameters in TB, suggests that dissociation does not necessarily imply a
conducting state. This may help explain why the TB conductivities found at
low density and pressure are much higher than the experimental
ones\cite{Leno97}. 

Recently Pfaffenzeller and Hohl\cite{Pfaf97} carried out Car-Parrinello
ab-initio molecular dynamics (AIMD) simulations on the metallic side of the
M-I transition and find an even higher ``dissociation fraction''.  Because
of the extremely short life-time of the pairs they term the metallic liquid
``monatomic'', even though it retains significant pairing correlations.
Their lowest density is slightly higher than the metallization density
calculated for the experiments ($D_m = 0.4 mol/cm^3$ vs. $0.32 mol/cm^3$)
and they find a resistivity about a factor 6 smaller than the experiments.
Simulating the metal-insulator transition with AIMD is difficult because the
local density approximation (LDA) underestimates the gap, and the
sensitivity on k-point sampling\cite{Edwa96} (the AIMD calculations above
use only one k-point) makes accurate large-scale simulations computationally
intensive. Once again, the large Hubbard U implies that spin correlations
may be important, necessitating AIMD simulations based on spin density
functional theory instead of the usual paramagnetic ansatz.

But whether any of these these mechanisms describe the shock
experiments\cite{Weir96} fully has yet to be established.  The gap closes
with increasing density even in an entirely diatomic phase, and if the TB or
AIMD even slightly over-estimate the ``dissociation'', the metallization in
the simulations will occur through the ``dissociation mechanism'' instead of
through a pure band-overlap scenario.  In fact, as evidenced from comparison
of pair correlation functions, the TB and AIMD results give a larger
``dissociation'' fraction than the path-integral monte-carlo (PIMC) results
of Magro {\em et al.}\cite{Magr96}.  While PIMC includes the substantial
zero-point motion of the protons which the TB and AIMD neglect, PIMC is
usually carried out in a much smaller cell.  Presently it is not clear which
method is best suited to the experimental regime and the sensitive
interdependence of dissociation and metallization physics coupled with a
possible role for strong correlations will render an unambiguous
interpretation of the simulations difficult.

Early predictions of finite-temperature pressure-induced metallization fall
under the general rubric of {\em Plasma Phase Transitions} (PPT)\cite{ppt}.
Most theories take a chemical approach, that is the species (bound states)
are well defined, and this is at variance with the extremely short life-times
predicted by simulations. Besides ignoring the possibility of band-overlap
metallization, the chemical models are typically based on pair-potentials and
a free-rotation approximation, both of which are ill-defined at these
densities as we have noted.  The most comprehensive chemical model
calculations were carried out by Saumon and Chabrier\cite{Saum92} who
predict for the PPT a critical point at $T=15310$ K and a density of $D_m =
0.18$ $mol/cm^3$, and at higher densities and lower temperatures an
increasingly discontinuous transition between a diatomic insulating state
and an ionized (conducting) monatomic phase (see also Fig.
(\ref{phasediagram})). In support of these chemical models, the physically
based PIMC calculations\cite{Magr96} also give rather similar results.  On
the other hand the TB simulations give a gradual transition to a more
``monatomic'' phase.  The metallization experiments appear just out of the
range of the PPT predictions.  To establish the existence of a PPT by shock
experiments will also require a range of temperature and density data.  

Clearly the experiments are in a very interesting regime; the interplay of
thermal and electronic effects brings the system close to both the closing
of the gap {\em and} the onset of pair fragmentation, the two being
interconnected.  Which occurs first will depend sensitively on the actual
temperature and pressure and one could imagine different behaviors at
different state points.  For example at densities just below the proposed
diatomic metal-insulator transition at 0 K one would expect a I-M transition
of the band-overlap type while at higher temperature and lower densities
structural change, if it exists, becomes more important.  A definitive
electronic interpretation of the Livermore shock experiments is still
lacking, and could prove to be quite unlike the conventional explanations of
transport properties of ordinary liquid metals.

\section{Future experimental probes}

Considerations based both on a simple band-overlap scenario within the Ziman
formulation, and the ``dissociation'' based mechanism proposed by Ross and
Lenosky {\em et al.}\cite{Leno97,Ross96}, predict that the conductivity is high, but
that it should drop
with increasing density.  Experiments going to higher pressure would
therefore be of great interest.  However, so also would shock experiments at
about the same compression but  lower temperatures.  If the arguments
given above are valid, then as depicted in Fig. (\ref{phasediagram}), a
boundary should be crossed (at fixed P) locating the I-M transition.  One
could also envisage that at even higher pressures, the I-M transition would
be at low enough temperature to be within reach of static pressure techniques
such as diamond anvil cells\cite{diamond}.

The transition to a conducting state in hydrogen may be quite unusual.  The
tenacity with which H retains its pairing, an effect driven largely by
exchange, can be manifested even in a metallic environment and at quite high
temperatures.  Though covalent, this ensuing state is not of a network
character as in molten Si (or Ge), but a rapid exchange of pairs (akin to
the fast exchange proceeding in Si and Ge) may be proceeding. It is an
interesting and perhaps novel state of matter, which now merits considerable
further study.

{\bf ACKNOWLEDGMENTS }

	We thank M. Ross and T. W. Barbee, III for discussions and
calculational results.  We acknowledge discussions with F. Hensel, P. P.
Edwards, T. J.  Lenosky, L. A. Collins, N. C. Holmes, and D. A. Young.  We
thank O. Pfaffenzeller and D. Hohl for sending their manuscript prior to
publication.  Work was performed under the auspices of the U.  S. Department
of Energy under Contract No.  W-7405-ENG-48 (WJN), and supported by the NSF
under Grant No.  DMR-24-8330 (AAL and NWA).

\begin{figure}
\caption[Fig1]{{\bf Schematic of electrical conductivity experiments on
fluid metallic hydrogen.}} Four electrodes in (a) were connected to the
circuit in (b).  For conductivities lower than metallic, two probes were
used.  Trigger pins turn on the recording system.  All cables are coaxial.
\end{figure}

\begin{figure}
\caption[Fig2]{{\bf Effect of rise time on pressure-density states. }} (a)
First pressure in hydrogen is $\sim P_f/30$, where $P_f$ is incident shock
pressure in $Al_2O_3$.  Successive reverberations comprise a quasi-isentrope
up to pressure $P_f$.  This quasi-isentrope is represented by ramp over
$\sim 50 ns$ from $P_f/30$ up to $P_f$ .  After reverberation is complete,
$P_f$ is held for $\sim 100 ns$.  If $P_f$ were achieved in one jump, this
state would be on single-shock Hugoniot.  (b) Equation-of-state curves
plotted as pressures versus density: 0 K isotherm, points reached by shock
reverberations, and single-shock Hugoniot.  Initial point is liquid $H_2$ at 1
atm.
\end{figure}

\begin{figure}
\caption[Fig3]{{\bf Logarithm of electrical resistivity of $H_2$ and $D_2$
plotted versus pressure.}} The slope change at 140 GPa is identified as the
transition from semiconducting to metallic fluid.
\end{figure}

\begin{figure}
\caption[FigE]{{\bf Schematic Phase Diagram}} The metal-insulator transition
in the liquid occurs at considerably lower pressure than the predicted
transition in the solid. We suggest a metal-insulator band starting from the
zero-temperature solid, and increasing in temperature with decreasing
density as depicted above.  At very low temperatures it will be sharp with
the character of a thermodynamic phase transition.  At the other extreme of
low density and very high temperatures, the metallization will be by gradual
transformation to a temperature ionized plasma. At zero temperature the
metal-insulator transition is predicted to occur before (at lower density
than) the diatomic-atomic transition.  What happens at higher temperatures
remains to be established.  Possibilities include the crossing or merging of
the lines, as well as their termination in one or two critical points.
Recent theoretical predictions of a Plasma Phase Transition
(PPT)\cite{Magr96,Saum92} are schematically depicted as well.
 Also depicted in the figure is the Simon equation for the melting line, and
the orientationally disordered (I) and ordered phases (II) and (III) found
in the solid\cite{Mao94}.  We emphasize that this phase diagram is merely
schematic; much still needs to be filled in.
\label{phasediagram}
\end{figure}


\begin{references}
\bibitem{Wign35} E. Wigner and H.B. Huntington, J. Chem Phys. {\bf 3}, 764
(1935).
\bibitem{Alde60} B. J. Alder and R. H Christian, Phys. Rev. Lett. {\bf 4},
450 (1960).
\bibitem{Chen96} H. N. Chen, E. Sterer, and I. F. Silvera, Phys. Rev. Lett.
{\bf 76}, 1663 (1996).
\bibitem{Heml96} R. J. Hemley, H. K. Mao, A. F. Goncharov, M. Hanfland, and
V.  Struzhkin, Phys. Rev. Lett. {\bf 76}, 1667 (1996).
\bibitem{Ruof96} A. L. Ruoff, in {\em High Pressure Science and Technology},
edited by W.  Trzeciakowski (World Scientific, Singapore, 1996), pp. 511-516.
\bibitem{Frie77} C. Friedli and N.W. Ashcroft, Phys. Rev. B {\bf 16}, 662
(1977).
\bibitem{Garc90} A Garcia, T. W. Barbee, M. L. Cohen, and I. F. Silvera,
Europhys. Lett.  {\bf 13}, 355 (1990).
\bibitem{Kaxi91} E. Kaxiras, J. Broughton and R.J. Hemley, Phys. Rev. Lett.
 {\bf 67}, 1138 (1991).
\bibitem{Chac92} H. Chacham, X. Zhu, and S.G. Louie, Phys. Rev. B {\bf 46},
6688 (1992).
\bibitem{Loub96} P. Loubeyre, R. LeToullec, D. Hausermann, M. Hanfland, R.
J. Hemley H. K. Mao, and L. W. Finger, Nature {\bf 383}, 702 (1996).
\bibitem{Ashc68} N.W. Ashcroft, Phys. Rev. Lett.  {\bf 21}, 1748 (1968).
\bibitem{Rich97} C. F. Richardson and N. W. Ashcroft, Phys. Rev. Lett. 
{\bf 78}, 118 (1997).
\bibitem{metallic} There is some ambiguity in the definition of a metal. One
possibility is the existence of a reasonably well defined Fermi-surface,
while in 1996, the late Sir Nevill Mott wrote: ``I've thought a lot about
`What is a metal' [and] I think one can only answer the question at $T=0$.
Thus a metal conducts [and] a non-metal doesn't.'' (P.P. Edwards, private
communication).  However, in this paper
 we use an operational definition:  A substance
is metallic if the conductivity is in a ``typical'' metallic range.
\bibitem{Weir96} S.T. Weir, A.C. Mitchell and W.J. Nellis, Phys. Rev Lett.
 {\bf 76}, 1860 (1996).
\bibitem{Diat85} V. Diatschenko, C. W. Chu, D. H. Liebenberg, D. A.
Young, M. Ross, and R.  L. Mills, Phys. Rev. B {\bf 32}, 381 (1985).
\bibitem{Nell92} W. J. Nellis, A. C. Mitchell, P. C. McCandless, D. J.
Erskine, and S.  T.  Weir, Phys. Rev. Lett. {\bf 68}, 2937 (1992).
\bibitem{Hawk78}   R.
S. Hawke, T. J. Burgess, D. E. Duerre, J. G. Huebel, R. N. Keeler, H.
Klapper, and W. C. Wallace, Phys. Rev. Lett. {\bf 41}, 994 (1978).
\bibitem{Nell83} W. J. Nellis, A. C. Mitchell, M. van. Thiel, G. J. Devine,
R. J.  Trainor, and N. Brown, J. Chem. Phys. {\bf 79}, 1480 (1983).
\bibitem{Holm95} N. C. Holmes, M. Ross, and W. J. Nellis, Phys. Rev. B {\bf
52}, 15835 (1995). 
\bibitem{Silv97} L. B. Da Silva, P. Celliers, G. W. Collins, K. S. Budil, N.
C. Holmes, T.W. Barbee, III, B. A. Hammel, J. D. Kilkenny, R. J. Wallace,
M.  Ross, and R.  Cauble, Phys. Rev. Lett. {\bf 78}, 483 (1997).
\bibitem{Butl96} R. P. Butler and G. W. Marcy, Astrophys. J. {\bf 464}, L153
	(1996); R. P. Butler and G. W. Marcy, Astrophys. J. {\bf 474}, L115
	(1996); M. Mayor and
D.  Queloz, Nature {\bf 378}, 355 (1995).
\bibitem{Zhar92} V. N. Zharkov and T. V. Gudkova, in {\em High-Pressure
Research:  Application to Earth and Planetary Sciences}, Y. Syono and M. H.
Manghnani, Eds., (Terra Scientific Publishing Co., Tokyo, 1992), pp.393-401.
\bibitem{Ross81} M. Ross, H. C. Graboske, and W. J. Nellis, Phil. Trans. R.
Soc. Lond.  A {\bf 303}, 303 (1981).
\bibitem{Stev83}     D. J. Stevenson, Rep. Prog. Phys. {\bf 46}, 555 (1983).
\bibitem{Nell95} W. J. Nellis, M. Ross, and N. C. Holmes, Science {\bf 269},
1249 (1995); W. J. Nellis, S. T. Weir, and A. C. Mitchell, Science {\bf
273}, 936 (1996).
\bibitem{Ashc95} N. W. Ashcroft, Physics World {\bf 8}, 43 (1995).
\bibitem{Edwa97} B. Edwards and N.W. Ashcroft, submitted (1997).
\bibitem{Jone66} A. H. Jones, W. M. Isbell, and C. J. Maiden, J. Appl. Phys.
{\bf 37}, 3493 (1966).
\bibitem{Mitc81a} A. C. Mitchell and W. J. Nellis, Rev. Sci. Instrum. {\bf
52}, 347 (1981).
\bibitem{Mitc81b} A. C. Mitchell and W. J. Nellis, J. Appl. Phys. {\bf 52},
3363 (1981).
\bibitem{Ersk94} D. Erskine, in {\em High-Pressure Science and
Technology-1993}, edited by S.  C.  Schmidt, J. W. Shaner, G. A. Samara, and
M. Ross (American Institute of Physics, New York, 1994), pp.141-143. 
\bibitem{Nell80} W. J. Nellis and A. C. Mitchell, J. Chem. Phys. {\bf 73},
6137 (1980).
\bibitem{Rado86} H. B. Radousky, W. J. Nellis, M. Ross, D. C. Hamilton, and
A. C.  Mitchell, Phys. Rev. Lett. {\bf 57}, 2419 (1986).
\bibitem{Yoo89} C. S. Yoo, G. E. Duvall, J. Furrer, and R. Granholm, J.
    Phys. Chem.  {\bf 93},3012 (1989).
\bibitem{Kerl83} G. I. Kerley, in {\em Molecular-Based Study of Fluids},
edited by J. M. Haile and G. A. Mansoori (American Chemical Society,
Washington, 1983), pp.  107-138.
\bibitem{Mott71a}       N. F. Mott, Phil. Mag. {\bf 24}, 2 (1971).
\bibitem{Hens96}  F. Hensel and P. P. Edwards, Phys. World, April, {\bf 43}
(1996).

\bibitem{Mott71b} N. F. Mott and E. A. Davis, Electronic Processes in
Non-Crystalline Materials (Oxford, London, 1971), p. 81.
\bibitem{Zima61} J. M. Ziman, Phil. Mag. {\bf 6}, 1013 (1961).
\bibitem{Fabe72} T.E. Faber {\em Introduction to the Theory of Liquid Metals}
 (Cambridge University Press, Cambridge 1972).
\bibitem{Shim77} M. Shimoji, {\em Liquid Metals} (Academic Press, London
1977).
\bibitem{resliquid} The resistivity of the liquid metallic state is often
easier to calculate than that of the solid metallic state because in the
former the scattering is dominated by density fluctuations while in the
latter the scattering may come from a multitude of different sources such
as defects, impurities etc...
\bibitem{Loui97} A.A. Louis, PhD thesis, Cornell Univeristy (1997).
\bibitem{struckcorr}
 In the usual application of the Ziman formalism, it is assumed that the
temperature scale is such that $k_BT >> \hbar \omega$ for all excitations of
interest in the system.  While these conditions hold for translational and
rotational energy scales, this is not true for vibrons.  In fact, the
temperature is quite closely matched to the vibron energies; a confluence
that leads to the possibility of near resonant scattering processes.  By
taking fixed diatomic molecules we ignore the effect of vibrational
excitation on the structure factor $S(k)$.  We use an expression for the
diatomic structure factor that includes the bond-length of $1.4 a_0$ and a
Percus-Yevick structure factor with $\sigma = 0.29$ and $\eta = 0.29$ for
the center-of-mass-center-of-mass part and as a first approximation neglects
contributions from
 anisotropy (i.e. the free-rotation approximation (see e.g. J-P. Hansen and
I. R. McDonald, {\em Theory of Simple Liquids}, (Academic Press, London
1986))). The rotational and translational time-scales are about equal at
this density, making the free-rotation approximation suspect. But even though
simulations\cite{Leno97,Pfaf97} give extremely short life-times for the
``molecules'', a structure factor based on the static pair and free-rotation
approximations does reasonably well (see for example A. Bunker, S. Nagel, R.
Redmer and G. R\"{o}pke,
   Phys. Rev. E (1997), to appear).
\bibitem{scattcorr}The neglected vibrations lead to inelastic scattering
effects.
Baym (G. Baym Phys. Rev {\bf 135}, A1691 (1964)) showed that in this case
the crucial quantitity is $S(\vec{k},\omega)$, the dynamic structure factor
for the protons defined by:

$S(\vec{k},\omega) = \int_{-\infty}^{\infty} dt e^{-i\omega t} <
\hat{\rho}(\vec{k},t)
\hat{\rho}^\dagger(\vec{k},0) >$

where $\hat{\rho}(k)$ is the one-particle density operator for protons) the
resistivity is then given by:

$\rho \sim \int_0^{2k_f} q^3 dq |v(q)|^2 \int_{-\infty}^{\infty} \frac{d
\omega}{2 \pi} S(\vec{k},\omega) \frac{\beta \hbar \omega}{(\exp(\beta \hbar
\omega) -1)} $

which reduces to the usual Ziman formula for $k_BT >> \hbar \omega$.  The
effect of the vibrations can be calculated and gives a correction to the
static structure factor $S(q)$ proportional to $q^2$, but with a small enough
prefactor (the same order as the correction for finite ion mass) to be
ignored at this level of approximation.
\bibitem{Stev74} D. J. Stevenson and N. W. Ashcroft, Phys. Rev. A {\bf 9},
782 (1974).
\bibitem{noEdwards} One reason for the success of the Ziman formalism is the
so-called Edwards cancellation theorem which states that the d.o.s.
corrections (as manifested in the effective mass) cancel out to first order
(see for example\cite{Fabe72,Shim77}).  However, in this case the Edwards
cancellation theorem probably doesn't hold (see e.g. M. Itoh and M. Watabe,
J. Phys. F {\bf 14}, L9 1984), and instead Mott's $g^2$ factor, where $g$ is
the ratio of the actual density of states to the free-electron value, must be
included\cite{Mott90}.  The minimum value of $g$ is typically about $0.3$,
raising the resistivity by a maximum of about $1/g^2 \sim  10$.
\bibitem{meanfreepath} The mean-free path is roughly given by: $l = [174
a.u.] \times [(rs/a_0)^2/\rho_\mu]$\cite{Ashc76}, which implies a mean free
path of about $9 a.u$. for a calculated resistivity of $\sim 50 \mu
\Omega cm$.  Thus the Ziman formalism is self-consistent in that sense.  A
lower effective carrier density raises $rs$, so that in this approach the
mean free path is still greater than the inter-molecular spacing for the
larger (experimental) resistivities.
\bibitem{Kita95} See for example H. Kitamura and S. Ichimaru, Phys.  Rev. E,
{\bf 51}, 5006 (1995).
\bibitem{Kres97} G. Kresse and J. Hafner, Phys. Rev. B, {\bf 55}, 7539 (1997).
\bibitem{Ashc95b} N.W. Ashcroft, {\em Elementary Processes in Dense Plasmas,}
 ed by S. Ichimaru and S. Ogata, (Addison Wesley, 1995).
\bibitem{Mao94} H-K. Mao and R.J. Hemley, Rev. Mod. Phys. {\bf 66}, 671
(1994).
\bibitem{Mott90} N. Mott, {\em Metal Insulator Transitions}
(Taylor \& Francis, London 1990).
\bibitem{Bismuth} This in contrast to some narrow gap semi-conductors, or
semi-metals like Bi, which would become a good metal upon a very slight
distortion of the structure (see e.g. \cite{Ashc76}).
\bibitem{bands}  The reader might wonder why we use  language
inspired by bands in this manifestly disordered system where the wave vector
${\bf k}$ is not at all a good quantum number.  Although the concept of energy
bands is not entirely correct here, it is well known that for many liquid
metals the band structure of a solid with similar local coordination gives a
good approximation of the electron physics of the liquid\cite{Fabe72}, and
so this is the sense in which we use the concept of bands here.
\bibitem{Ashc76} N.W. Ashcroft and N.D. Mermin, {\em Solid State Physics}
(Holt, Rinehart and Winston, New York 1976).
\bibitem{Halp67}See e.g. B.I. Halperin and
T.M.  Rice, Solid State Physics {\bf 21}, 116 (1968).
\bibitem{Thou79} D. J. Thouless in {\em Ill-Condensed Matter} edited by
R. Balian, R. Maynard and G. Toulouse (North Holland, Amsterdam 1979).
\bibitem{Edwa83} P.P. Edwards and M. Sienko, Chemistry in Britain {\bf 19},
39 (1983), M. Ross, J. Chem. Phys {\bf 56}, 4651 (1972).
\bibitem{WignerMott} Its interesting that at their original predicted
density, the monatomic solid would be close to a Mott-Hubbard insulator!
\bibitem{well} Full  ab initio LDA calculations of the pair potential
in the solid phase at the liquid metallization densities give a well depth
of about $3$ eV, and also the correct vibron frequency of about $0.5$ eV.
These calculations also show very strong many-body effects making
pair-potentials suspect ( see for example, B. Edwards and N.W. Ashcroft,
unpublished). 
 Even free $H_2^+$ is
strongly paired  with a dissociation energy of about $2.8$eV, a fact which
helped motivate an earlier prediction of metallization in the molecular
liquid\cite{Ashc68}. In the solid, vibron overtones have been measured up
to pressures of $170$ GPa\cite{Chen96,Heml96}.
\bibitem{Leno97} T.J. Lenosky J.D. Kress, L.A. Collins, I Kwon, Phys. Rev. B
{\bf 55}, 11907 (1997). T.J. Lenosky J.D. Kress, L.A. Collins, I Kwon J. Quant.
Spect. Radiat. Trans (1997), to appear.
\bibitem{Ross96} M. Ross, Phys. Rev. B {\bf 54}, R9589 (1996).
\bibitem{Pfaf97}  O. Pfaffenzeller and D. Hohl, Preprint 1997.
\bibitem{Edwa96} See for example B. Edwards, N.W. Ashcroft and T.J. Lenosky,
Euro. Phys.  Lett. {\bf 34}, 519 (1996) and J. Kohanoff {\em et al.} Phys. Rev.
Lett. {\bf 78}, 2783 (1997).
\bibitem{Magr96} W.R. Magro, D.M. Ceperley, C. Pierleoni and B. Bernu,
Phys. Rev. Lett. {\bf 76}, 1860 (1995), W.R. Magro, PhD thesis, University
of Illinois (1995).
\bibitem{ppt} The PPT was first proposed by L. Landau and G. Zeldovich, Acta.
Phys. Chim. USSR, {\bf 18}, 194 (1943) for a somewhat different system.
Since then the term PPT has been applied to a multitude of different
transitions of the insulating-conducting type.
\bibitem{Saum92} D. Saumon and G. Chabrier, Phys. Rev. A {\bf 46}, 2084
(1992).
\bibitem{diamond} Heating of hydrogen can cause diamond anvils to fail, but
at high enough pressures the the metal-insulator transition should occur at
temperatures that may be low enough to avoid this failure.

\end{references}
\end{document}